\documentstyle[11pt]{article}
\date{}
\title{Towards the mass spectrum of quantum black holes and wormholes}
\author{V.A.Berezin${}^1$,A.M.Boyarsky${}^2$,A.Yu.Neronov${}^2$\\
${}^1${\it Institute for Nuclear Research of the Russian Academy of Sciences,
Moscow, Russia}\\
${}^2${\it Dept.Mathematics and Mechanics, Lomonosov Moscow State Univ.,
Moscow, Russia}}
\begin{document}
\maketitle
\begin{abstract}
The quantum black hole model with a self-gravitating spherically symmetric 
thin dust shell as a source is considered. The shell Hamiltonian constraint 
is written and the corresponding Schroedinger equation is obtained. 
This equation appeared to be a finite differences equation. Its solutions 
are required to be analytic functions on the relevant Riemannian surface. 
The method of finding discrete spectra is suggested based on the analytic 
properties of the solutions. The large black hole approximation is considered 
and the discrete spectra for bound states of quantum black holes and wormholes 
are found. They depend on two quantum numbers and are, in fact, 
quasi-continuous. The quantum black hole bound state depends not only on 
mass but also on additional quantum number, and black holes with the same mass 
have different quantum hairs. These hairs exhibit themselves at the 
Planckian distances near the black hole horizon. For the observer who can not 
measure the distances smaller than the Planckian length the black hole has 
the only parameter, its mass. The other, non-measurable parameter leads to 
existence of the black hole entropy.
\end{abstract}

This paper can be considered as a short version of the paper \cite{we} with 
the emphasis on the mass spectrum of quantum black holes and wormholes. 
Thus, we pay attention only to the crucial points that lead to the derivation 
of these spectra.

1. We start with description of our model. It is a self-gravitating spherically 
symmetric thin dust shell endowed with mare mass $M$. Note that the shell is 
not embedded into the Schwarzschild manifold in which case it can be 
considered as some set of test particles (observers). Our shell serves as a 
source of a gravitational field. Inside the shell the space-time is 
Minkowskian, and outside it is Schwarzschildean with mass $m$. In what follows 
we need some well known facts from the classical theory of black holes. Every 
spherically symmetric space-time can be locally characterized by two 
invariant functions of two variables (some time coordinate $t$ and some 
radial coordinate $r$). These are the radius of a sphere $R(t,r)$ and the 
differential invariant $F(t,r) = g^{\alpha,\beta}R_{,\alpha}R_{,\beta}$, the 
latter being equal to $F = 1 - 2Gm/R$ in the Schwarzschild case. The complete 
Schwarzschild manifold consists of four parts characterized by the signs of 
function $F$ and the signs of partial derivatives of $R(t,r)$, they are 
called $R_{\pm}$- and $T_{\pm}$-regions. In the $R$-regions $F > 0$, and 
$R^{\prime} > 0$ in $R_+$-region ($R$ ranges between the event horizon 
$R_g = 2Gm$ and infinity) and $R^{\prime} < 0$ in $R_-$-region 
($\infty > R > R_g$). In the $T$-regions $F < 0$, and the $T_+$-region in 
which $\dot R > 0$ is called the region of inevitable expansion while the 
$T_-$-region with $\dot R < 0$ is called the region of inevitable contraction. 
We are interested here in the bound motion only. So, a trajectory of our 
shell has a turning point of radius $R_0$ which can be located in one of the 
$R$-regions (but not in the $T$-regions). It appears that if the ratio of 
the total (Schwarzschild) mass $m$ and the bare mass $M$ is in the range 
$1/2 < m/M < 1$, then the turning point lies in the $R_+$-region, we call 
this a black hole case. If $m/M < 1/2$, then the turning point is in the 
$R_-$-region and we call this a wormhole case. The ``black hole'' shell 
does not enter the $R_-$-region, and the ``wormhole'' shell does not enter 
the $R_+$-region. The main feature of the ``black hole shells is that for 
fixed $R_0$ the larger the bare mass, the larger the total mass, i.e. 
$\frac{\partial m}{\partial M} > 0$. For the ``wormhole'' shells 
$\frac{\partial m}{\partial M} < 0$.

2.To construct a quantum theory of black holes and wormholes we need a 
classical geometrodynamical description of our model. The geometrodynamics 
of the eternal Schwarzschild black hole (both classical and quantum) was 
considered in full details by K.Kuchar \cite{kuchar}. The geometrodynamics 
of the general spherically symmetric space-time with the thin shell as a 
source was constructed in our paper \cite{we}. It was shown that the 
corresponding Hamiltonian constraint for the shell depends only on the 
invariant functions $R$ and $F$, on the bare mass of the shell $M$ and the 
momentum $P_R$ conjugate to the variable $R$. It can be written in the form

\begin{equation} 
\label{Class}
C = F + 1 - \sqrt{F}\left( \exp\frac {G P_R} {R}+
 \exp - \frac {G P_R} {R}\right) - \frac{M^2G^2}{R^2} = 0
\end{equation}
Strictly speaking, the above equation was derived for $R_+$-region only. 
Because $\sqrt{F}$ it is not valid as it is in $T$-regions. Of course, the 
analogous equations can also be derived separately for $T$-regions. But, 
having in mind that in quantum theory it is desirable to have a single wave 
function for all the four patches of the complete Schwarzschild manifold 
we have chosen quite a different way. We consider $f = \sqrt{F}$ as a function 
complex variable, namely, $f = |F|^{1/2}e^{i\phi}$, which has branching 
points at the horizons, when $F = 0$. We choose the following rules of 
bypassing around these branching points. In the black hole case $\phi = 0$ 
in the $R_+$-region, $\phi = \frac{\pi}{2}$ in the $T_-$-region, 
$\phi = \pi$ in the $R_-$-region and $\phi = -\frac{\pi}{2}$ in the 
$T_+$-region. In the wormhole case the bypass goes in the opposite direction 
starting from the $R_-$-region. The Hamiltonian constraint is now a complex 
valued function but it can easily be made real by adding the relevant complex 
conjugate part. The advantage of such an analytical continuation is not only 
that we can now obtain a single wave function for a quantum self-gravitating 
shell, but what is more important the $R_+$- and $R_-$-regions of the 
complete Schwarzschild manifold are no more identical but can be considered 
as lying on different leaves of Riemannian surface of complex variable 
$f (= |F|^{1/2}e^{i\phi})$. We will see soon that this fact affects the 
quantum mass spectrum very much. We are not going to consider here the 
classical evolution which comes from the above Hamiltonian, but will jump 
directly to the quantum picture.

3.In the quantum theory both the variables and their conjugate momenta
become operators, and the Hamiltonian constraint acts on the wave function as 
operator. In our case it is more convenient to use the radius $R$ and its 
conjugated momentum $P_R$ but the equivalent canonical pair $s=R^2/R_g^2$
and $P_s=\frac{P_g^2}{2R}P_R$. In the coordinate representation the wave
function $\Psi$ depends only on s, which is a multiplication operator ,
and $P_s$ becomes a differential operator $P_s=-i\frac{\partial}{\partial s}$.
The exponential operator $exp(GP_R/R)=exp(-i\xi\frac{\partial}{\partial s})$  
which enters our Hamiltonian constraint becomes in the coordinate representation
an operator of finite shift.   

\begin{equation}
\label{finite}
e^{-i\xi\frac{\partial}{\partial  s}}\Psi =
\Psi( s -\xi i)
\end{equation}
where $m_{pl}$ is Plank mass and $\xi =\frac {1}{2} (\frac {m_{pl}}{ M})^2 $ .
Thus , we arrive at the following quantum equation for a self gravitating thin
 dust shell

\begin{equation}
\label{main'}
\begin{array}{c}
f\left( \Psi (s+ i\xi)+\Psi (s- i\xi)\right)+
\overline{f}(s+ i\xi)\Psi (s+ i\xi)+\\
\overline{f}(s- i\xi)\Psi (s- i\xi)=
2(F+1-\frac{1}{\displaystyle {4s}})\Psi(s)
\end{array}
\end{equation}
where $f=|F|^1/2e^{i\phi}$ and we have chosen a symmetric operator ordering
with an appropriate complex conjugation. The quantum equation obtained is the
equation  in finite differences rather than the differential one as in ordinary
quantum mechanics, and the shift is along imaginary axis. The quantum 
mechanical postulates tell us that the Hamiltonian (as the constraints as well)
should be self-adjoint operators. And this goal is achieved usually by imposing
appropriate boundary conditions on the wave functions. As by product, for bound
states we obtain usually a discrete energy (mass) spectrum. The reason for this
is that for any homogeneous ordinary differential equation , say, of second
 order we need only one condition to single out the solution (up to the 
renormalization factor). But for the corresponding quantum operator to be a 
self-adjoint we need two boundary conditions for bound states . It is such 
an extra condition which leads to the discrete spectrum . The situation
is different in the case of our equation in finite differences . In the paper 
\cite{shell} the toy quantum black hole model was considered. The  quantum 
equation in this model is also equation in the finite differences. This 
equation appeared to be exactly solvable and it was possible shown that it was
 possible to obtain a self-adjoint extension of the of the corresponding 
Hamiltonian by  imposing of the countable number of boundary conditions on the
wave functions. All these boundary conditions allowed to single out the 
solution (up to the inherent degeneracy), but they did not allow to obtain the
discrete spectrum. We expect  our problem to have  a discrete  mass (energy) 
spectrum for bound states because the same problem in 
the nonrelativistic limit has this feature. How to obtain it ?
Fortunately , we have one more requirement the wave function should satisfy.
The solution to the second order differential  equation should be 
differentiable twice (at least). But now we have at hand an equation in finite 
differences. Moreover the shift is along the imaginary axis. Therefore , we 
must require that the solution should be analytical function. The analyticity
is a very strong feature. Our equation has singular points, in particular, the
 points at the horizons (s=1) are the branching points. The solutions , 
in general , will have branching points  too. And the types of these branching 
points do not depend neither on the particular operator ordering nor on the 
boundary conditions imposed on the wave functions to ensure the 
self-adjointness of th Hamiltonian (but the wave function should still be 
integrable with square). Moreover,in order to be able to construct the wave 
function which is single-valued on some Riemannian surface, we should have the
 branching points  of the same type (when this points can be connected by the 
cuts). It is the comparison of the branching points of the ``good'' (integrable)
solutions that will lead us to the  discrete spectra for bound states. We will
see in a moment how such a procedure works in the limit of large black holes
(e.i., large total mass m.).

4.We would like to illustrate the ideas described above by considering the 
limiting case of small values of $\xi$. Since 
$\xi = \frac{1}{2}(m_{pl}/m)^2$, it is not only the limit of black holes with 
masses much larger than the Planckian mass, but at the same time it is a 
quasiclassical limit because $m_{pl}^2$ is proportional to the Planckian 
constant $\hbar$. In this limiting case we can expand any function of 
shifted argument in the following series

\begin{equation}
\Phi (s+\xi i)=\Phi (s)+i\xi \Phi'(s)-
\frac{\displaystyle \xi^2}{\displaystyle 2}\Phi''(s)+...  
\end{equation}
We cut these series at the second order in $\xi$. Here we present only the 
results of our investigation. The resulting equations are different for 
different parts of the Schwarzschild manifold but all of them have the same 
singular points as the original equation. These points are 
$s \rightarrow \infty$, $s \rightarrow 1+0$ in $R_{\pm}$-regions, and 
$s \rightarrow 1-0$ in $T_{\pm}$-regions (our expansion is not valid near 
$s = 0$ and this point is not relevant to the results). As was explained 
before we need only to know the asymptotic behavior of the solutions near 
singular points of the equations. Below we consider the black hole case only. 
The results are readily translated to the wormhole case. The very interesting 
feature of our equation is the fact that the approximate differential 
equations in $R_{\pm}$-regions are of the second order, while in 
$T_{\pm}$-regions the equations appear to be of the first order ones. 
The corresponding asymptotics at $s \rightarrow 1-0$ are
   
\begin {equation}
\Psi \sim 
\exp\left(i\frac{8}{\displaystyle 3\xi^2}
\left( 1-\frac{\displaystyle M^2}{\displaystyle 4m^2}\right)(-y)^{3/2}\right)
\end{equation}
in the $T_-$-region, and

\begin {equation}
\Psi \sim 
\exp\left(-i\frac{8}{\displaystyle 3\xi^2}
\left( 1-\frac{\displaystyle M^2}{\displaystyle 4m^2}\right)(-y)^{3/2}\right)
\end{equation}
in the $T_+$-region. The variables $s$ and $y$ are related by $s = (1 + y)^2$. 
Thus, $y$ is the deviation from the horizon. We see that the wave function 
is the radial part of the ingoing wave in the $T_-$-region and it is that of 
the outgoing wave in the $T_+$-region. This is a quasiclassical reflection of 
the fact that classically the shell can only expand in the $T_+$-region and 
shrink in the $T_-$-region. Moreover it shows that our choice of the 
quantum Hamiltonian constraint (operator ordering, complex conjugation etc.) 
gives a good quasi-classics. Let us remind that the solution to the original 
equation in finite differences should be an analytical function. The 
solution to the approximate equation should not have, of course, this 
feature. But the asymptotic solution on one side of the branching point 
should be the analytical continuation of the solution on the other side. This 
dictates the choice of one of the two asymptotics in the $R_+$-region  

\begin {eqnarray}
\Psi \sim 
1-\frac{8}{\displaystyle 3\xi^2}
\left( 1 - \frac{\displaystyle M^2}{\displaystyle 4m^2}\right)y^{3/2}\\
\nonumber
y\gg\xi,\ \ \xi\ll 1
\end{eqnarray}
and in the $R_-$-region
 
\begin {eqnarray}
\Psi \sim 
1-\frac{8}{\displaystyle 3\xi^2}
\left( 1 + \frac{\displaystyle M^2}{\displaystyle 4m^2}\right)y^{3/2}\\
\nonumber
y\gg\xi,\ \ \xi\ll 1
\end{eqnarray}
And, at last, the asymptotics at $s \rightarrow \infty$ in the $R_+$-region 
is 

\begin{eqnarray}
\Psi &\sim &
y^{\displaystyle \frac{1}{2}-\frac
{\displaystyle \frac{\displaystyle M^2}{\displaystyle m^2}-2}
{\displaystyle 4\mu\xi^2}}
\exp(-\mu y), \\
\nonumber
\mu&=&\frac{1}{\xi} 
\sqrt{\frac{\displaystyle M^2}{\displaystyle m^2}-1},
\ y\gg\xi
\end{eqnarray} 
while in the $R_-$-region it is 

\begin{equation}
\Psi \sim 
y^{\displaystyle \frac
{\displaystyle \frac{\displaystyle M^2}{\displaystyle m^2}-1}
{\displaystyle 8 \xi}}
\exp (-\frac {2}\xi y^2)
\end {equation}
Note that the falloff in the $R_-$-region is much more fast than in the 
$R_+$-region.

5. The last step on the way to the discrete mass spectrum for bound states 
of black holes and wormholes is to compare different branching points of 
solutions, namely, at infinities and near the horizons. But before doing this 
we would like to discuss some new and important point. Classically, given 
some total mass $m$, we have two types of motion depending on the value of 
mare mass $M$. In the first, black hole, case the bound motion starts from 
the past singularity $R = 0$ in the $T_+$-region, has its turning point in 
the $R_+$-region and then goes to the future singularity $R = 0$ in the 
$T_-$-region. The bare mass is such that $1 > \frac{m}{M} > \frac{1}{2}$, 
and $\frac{\partial m}{\partial M} > 0$. In the second, wormhole, case the 
turning point lies in the $R_-$-region with $\frac{m}{M} < \frac{1}{2}$, 
and $\frac{\partial m}{\partial M} < 0$. Quantum theory changes the situation 
radically. As we have seen the wave function in the black hole case is not 
zero not only in the $R_+$-region, but also in the $R_-$-region, though 
with relatively negligible amplitude. It means that the black hole type shell, 
starting from the $T_+$-region, may go not only through the ``true''  
$R_+$-region, but also through the ``wrong'' $R_-$-region (the same is true, 
of course, for the wormhole type with interchange of ``true''  and ``wrong'' 
regions). Since, by our construction, $R_+$-region and $R_-$-region lie on 
different leaves of the Riemannian surface. It means that the quantum shells 
have two degrees of freedom (contrast to the one degree for the classical 
shells). Consequently, the discrete mass spectrum should depend on two 
quantum numbers. And, indeed, the comparison of branching points of the 
solutions at $s \rightarrow \infty$ and $s \rightarrow 1+0$ in the 
$R_+$-region gives us the first quantum number,   
 
\begin{equation}
\label{n}
\frac
{\displaystyle 2-\frac{\displaystyle M^2}{\displaystyle m^2}}
{\displaystyle 4\zeta\sqrt{\frac{\displaystyle M^2}{\displaystyle m^2}-1}}
=n, \ \ n=integer
\end{equation}
Doing the same in the $R_-$-region we obtain the second quantum number, 

\begin{equation}
\label{p}
{\displaystyle \frac
{\displaystyle \frac{\displaystyle M^2}{\displaystyle m^2}-1}
{\displaystyle 8 \zeta}}=\frac{1}2+p, \qquad p=positive \quad integer
 \end {equation}
It can be shown that positive values of $n$ corresponding to the ``black hole''
shell with $\frac {\partial m}{\partial M}>0$, $1<\frac {M^2}{m^2}<2$ 
(instead of 4 for classical shells ), while negative values of 
$n$ correspond to the ``wormhole'' $\frac {\partial m}{\partial M}<0$,
 $\frac {\partial m}{\partial M}>4.2$ (instead of 4). Of course,
these numbers should be taken seriously, because we found the spectrum only
for large masses. Nevertheless, we see that there is a gap between range 
of masses for black holes and wormhole. 
We solved for $m$ the above quantization conditions in some limiting cases.

(i) $|q|=\frac{1+2p}{|n|\gg 1}$

\begin{equation}
\label{spectr1}
m\approx
\sqrt{2}\sqrt{1+2p}\  m_{pl}
\end{equation}
for black holes (n>0), and 

\begin{equation}
\label{spectr2}
m\approx
\sqrt{2}(\frac{|n|}{1+2p})^{1/2}\sqrt{|n|}\  m_{pl}
\end{equation}
for wormholes (n<0).
(ii) $0<q \ll 1$
\begin{equation}
\label{spectr3}
m\approx
\sqrt{2}(\frac{1+2p}{n})^{1/6}\sqrt{n}\  m_{pl}
\end{equation}
for black holes . There is no wormhole solution for small $|q|$, they exist only
 if $|q|\geq  \frac{3\ sqrt{3}}{2}$. If $|q|= \frac{3\ sqrt{3}}{2}$, than
$\frac{M^2}{m^2}\approx 4$, and 
\begin{equation}
m=\frac{2}{3^{1/4}}\sqrt{|n|}\  m_{pl}
\end{equation}

6. The appearance of the second quantum number means that the mass spectrum 
of quantum black holes is, in fact, quasi-continuous. It means also that the 
mass is not the only parameter that describes quantum black hole states. 
There exists quantum hairs, different for different black holes of the 
same mass. But these hairs exist on the Planckian distances from the black 
hole horizon. Let us assume that some observer can not measure distances 
smaller that the Planckian length (``natural coarse graining''). Then, for 
such an observer, a black hole is characterized by only one parameter, its 
mass. And this is just the origin of the black hole entropy.

We are greatly indebted to the Russian Foundation for Basic Researches
for the financial support (Grant N.97-02-17-064).
One of us (A.B.) thanks International Soros Science Education Program for
financial support.

\end{document}